\documentclass[pra,twocolumn,superscriptaddress,showpacs,amsmath,amssymb]{revtex4-1}
\usepackage{epsfig}
\usepackage{subfigure}
\usepackage{amssymb,amsmath}
\usepackage{graphicx}
\usepackage{epstopdf}
\usepackage{epsfig}
\usepackage{color}
\usepackage{amsfonts}
\usepackage{amscd}
\usepackage{float}
\usepackage[colorlinks,urlcolor=blue,linkcolor=blue,citecolor=blue]{hyperref}
\usepackage{bbm}
\usepackage{mathrsfs}
\usepackage{mathtools}
\emergencystretch=\maxdimen
\DeclareMathSizes{10}{10}{7}{5}

\begin{document}

\title{Generating and detecting entangled cat states in dissipatively coupled degenerate optical parametric oscillators}

\author{Zheng-Yang Zhou}
\affiliation{Beijing Computational Science Research Center, Beijing
100094, China}
\affiliation{Theoretical Quantum Physics Laboratory, RIKEN Cluster for Pioneering Research, Wako-shi, Saitama 351-0198, Japan}

\author{Clemens Gneiting}
\affiliation{Theoretical Quantum Physics Laboratory, RIKEN Cluster for Pioneering Research, Wako-shi, Saitama 351-0198, Japan}

\author{J. Q. You}
\altaffiliation[jqyou@zju.edu.cn]{}
\affiliation{Interdisciplinary Center of Quantum Information, State Key Laboratory of Modern Optical Instrumentation, and Zhejiang Province Key Laboratory of Quantum Technology and Device, Department of Physics, Zhejiang University, Hangzhou 310027, China}
\affiliation{Beijing Computational Science Research Center, Beijing
100094, China}
\author{Franco Nori}
\altaffiliation[fnori@riken.jp]{}
\affiliation{Theoretical Quantum Physics Laboratory, RIKEN Cluster for Pioneering Research, Wako-shi, Saitama 351-0198, Japan}
\affiliation{RIKEN Center for Quantum Computing (RQC), 2-1 Hirosawa, Wako-shi, Saitama 351-0198, Japan}
\affiliation{Physics Department, The University of Michigan, Ann Arbor, Michigan 48109-1040, USA}

\date{\today}

\begin{abstract}
Non-Gaussian continuous variable states play a central role both in the foundations of quantum theory and for emergent quantum technologies. In particular, ``cat states'', i.e., two-component macroscopic quantum superpositions, embody quantum coherence in an accessible way and can be harnessed for fundamental tests and quantum information tasks alike. Degenerate optical parametric oscillators can naturally produce single-mode cat states and thus represent a promising platform for their realization and harnessing. We show that a dissipative coupling between degenerate optical parametric oscillators extends this to two-mode entangled cat states, i.e., two-mode entangled cat states are naturally produced under such dissipative coupling. While overcoming single-photon loss still represents a major challenge towards the realization of sufficiently pure single-mode cat states in degenerate optical parametric oscillators, we show that the generation of two-mode entangled cat states under such dissipative coupling can then be achieved without additional hurdles. We numerically explore the parameter regime for the successful generation of transient two-mode entangled cat states in two dissipatively coupled degenerate optical parametric oscillators. To certify the cat-state entanglement, we employ a tailored, variance-based entanglement criterion, which can robustly detect cat-state entanglement under realistic conditions.
\end{abstract}

\maketitle

%
%

\section{introduction}
When Schr\"odinger imagined a cat in a quantum superposition of simultaneously being dead and alive, he intended to illustrate the seemingly absurd consequences when taking quantum mechanics too literally. Since then, however, macroscopically distinct quantum superpositions, ``Schr\"odinger cat states''~\cite{introcat1,introcat2,introcat3,introcat4,introcat5,introcat6}, and in particular entangled versions~\cite{ecs1, ecs2, ecs3, ecs4, ecs5,ecs6} thereof, have been identified as versatile resources, not only to explore the applicability of quantum mechanics in the macroscopic realm, but also to be utilized in quantum technologies. Consequently, great efforts have been invested to realize entangled cat states, and today, these are readily available on various experimental platforms \cite{expecs1, expecs2, expecs3, expecs4, expecs5, expecs6}.

Degenerate optical parametric oscillators (DOPOs)~\cite{DPO1,DPO2,DPO3,DPO4} have found widespread application in quantum optics, and their dynamics naturally comprises cat-like superpositions of coherent states in the limit of small single-photon loss. In addition, the generation of cat states and entangled cat states in DOPOs is of great interest for recent endeavors to deploy a {\it coherent Ising machine} (CIM)~\cite{CIM1, CIM2, CIM3, CIM4, CIM5, CIM6, CIM7, CIM8, CIM9}. The CIM represents a time-multiplexed network of DOPOs, with target applications in optimization, quantum simulation, and quantum information processing. Such DOPO networks excel in their flexibility to tailor individual couplings between arbitrary nodes. Cat states and entangled cat states could serve to encode, process, and read out discrete (quantum) information in these continuous variable systems. However, the successful production of cat states in DOPOs is, due to the difficulty in suppressing the single-photon loss, still challenging. Proposed strategies to overcome this decoherence include, for instance, the utilization of squeezed-state inputs \cite{CIM9}.

It is therefore important to understand under what conditions single-mode cat states and two-mode entangled cat states can be produced in DOPOs. In the case of single-mode cat states \cite{cat1, cat2, cat3, cat4, cat5, cat6, cat7, cat8, cat8}, a growing body of literature has studied potential ways, and detailed parameter regimes, towards their generation in DOPOs \cite{CIM9, dpocat1, dpocat2, dpocat3, dpocat4, dpocat5,dpocat6,dpocat7}. These results for single-mode cat states can potentially also be relevant to understand the generation of entangled cat states. For example, a dissipative coupling \cite{CIM2, CIM8} between two DOPOs, i.e., a dissipation process that acts collectively on both DOPOs, can drive a product state of two single-mode cat states into an entangled cat state, by suppressing a certain parity in the product state. Such dissipative couplings emerge, for instance, from the coupling of DOPOs via lossy delay lines in time-multiplexed DOPO networks. As this dissipative coupling can be strong compared to the single-photon leakage in DOPOs, this suggests that the main challenge for the generation of entangled cat states remains the production of single-mode cat states of sufficiently high quality; i.e., given single-mode cat states can be generated with high quality, the possibility to generate entangled cat states under the dissipative coupling follows.

In this work, we analyze in detail the generation of two-mode entangled cat states in two dissipatively coupled DOPOs. To this end, we first specify the theoretical model and justify in some detail the generic expectation that entangled cat states can be produced in this setup. Then we recapitulate the entanglement criterion applied in the article, which, by construction, detects entanglement in two-mode entangled cat states~\cite{mec1,mec2,mec3,mec4,mec5}. Subsequently, we determine the {\it threshold parameters under which entangled cat states (or states that are sufficiently close to these) are detectably present in the system}. In order to relate our threshold parameters to the generation of single-mode cat states, we also determine the fidelity of the cat state production under these threshold parameters when the dissipative coupling is switched off. Moreover, we investigate how the recently proposed environmental engineering~\cite{CIM9} towards enhanced cat state production carries over to the generation of entangled cat states.

In Sec.~\ref{theory}, we theoretically analyze the generation of two-mode entangled cat states and the detrimental effects of the single-photon loss. In Sec.~\ref{criterion}, we briefly introduce and discuss the entanglement criterion used here. In Sec.~\ref{parareg}, we determine the threshold parameters for generating detectable two-mode entangled cat states and compare these with the single-mode case. Section~\ref{bathsq} discusses possible improvements with environmental engineering. Section~\ref{conclusions} contains our conclusions.

\section{Theoretical model}\label{theory}

We consider two dissipatively coupled DOPOs. Including the effects of the pump modes up to second order and changing to the interacting picture, the system Hamiltonian can be expressed as (we set $\hbar=1$),
\begin{eqnarray}
H=\sum_{k=1}^2-iS[(a^{\dag}_k)^2-(a_k)^2],\label{systemhamiltonian}
\end{eqnarray}
with $S$ the effective pump intensity, and the annihilation (creation) operator $a_{k}$ ($a_{k}^{\dag}$) of the $k$th DOPO mode. The overall system evolution is governed by the quantum master equation
\begin{eqnarray}
\frac{d}{dt}\rho(t)&=&-i[H,\rho(t)]+\mathcal{L}_{\rm tot}[\rho(t)],
\end{eqnarray}
where the Lindblad superoperator $\mathcal{L}_{\rm tot}$ adds three dissipative contributions,
\begin{eqnarray} \label{Eq:dissipative_contributions}
\mathcal{L}_{\rm tot}(\rho)&=&\mathcal{L}_{\rm s}(\rho)+\mathcal{L}_{\rm d}(\rho)+\mathcal{L}_{\rm c}(\rho).
\end{eqnarray}
To express these dissipative terms more conveniently, we define
\begin{eqnarray}
\mathcal{L}[\Gamma,L,\rho(t)]&=&\frac{\Gamma}{2}\left[2L\rho(t)L^{\dag}-\left\{L^{\dag}L,\rho(t)\right\}\right],
\end{eqnarray}
with the dissipation rate $\Gamma$, some operator $L$, and the anti-commutator $\{A,B\}\equiv AB+BA$.
The first two terms in (\ref{Eq:dissipative_contributions}) describe the dissipation of single DOPOs, and are also relevant in the dissipative generation of cat states on other platforms \cite{cat6,cat8}.
\begin{eqnarray} \label{singleDPOH}
\mathcal{L}_{\rm s}(\rho)&=&\sum_{k=1}^2\mathcal{L}[\gamma_{\rm s},a_k,\rho(t)],\nonumber\\
\mathcal{L}_{\rm d}(\rho)&=&\sum_{k=1}^2\mathcal{L}[\gamma_{\rm d},a_k^2,\rho(t)].
\end{eqnarray}
Here, $\mathcal{L}_{\rm s}$ is the single-photon loss of each mode with rate $\gamma_{\rm s}$. The two-photon loss~\cite{dphotondecay1,dphotondecay2,dphotondecay3,dphotondecay4,dphotondecay5,dphotondecay6,dphotondecay7} $\mathcal{L}_{\rm d}$ is induced by the nonlinear coupling to a pump mode with strong dissipation. In Eq.~(\ref{singleDPOH}), $\gamma_{\rm d}$ is the effective two-photon dissipation rate. The third term in (\ref{Eq:dissipative_contributions}) describes the dissipative coupling between the two DOPOs,
\begin{eqnarray} \label{couplingH}
\mathcal{L}_{\rm c}(\rho)&=&\mathcal{L}[\gamma_{\rm c},a_1-a_2,\rho(t)].
\end{eqnarray}
The collective dissipation rate is $\gamma_{\rm c}$. Such a dissipative coupling, which has the power to generate entanglement, can be realized by a lossy mode coupled to both DOPOs. In time-multiplexed networks of DOPOs, such coupling emerges when two DOPOs are connected through a lossy delay line \cite{CIM2}.

\subsection{Ideal entangled-cat state generation}

We briefly discuss the mechanism underlying the generation of entangled cat states. If we consider only the system Hamiltonian~(\ref{systemhamiltonian}) and the two-photon dissipation $\mathcal{L}_{\rm d}$ in Eq.~(\ref{singleDPOH}), the resulting system assumes, for an initial vacuum state, the steady state
\begin{eqnarray}
\rho_{\rm steady}(t)&=&\frac{1}{2+\epsilon_{\rm sc}}{(|\alpha\rangle+|-\alpha\rangle)(\langle\alpha|+\langle-\alpha|)},\label{singlemodesteadystate}
\end{eqnarray}
with the coherent state amplitude $\alpha=i\sqrt{2S/\gamma_{\rm d}}$ and the normalization factor $(2+\epsilon_{\rm sc})$. Note that $\epsilon_{\rm sc}$ is a small correction depending on the overlap between $|\alpha\rangle$ and $|-\alpha\rangle$. The subindex ``sc" refers to single-mode cat state. The steady state $\rho_{\rm steady}$ is a cat state if $|\langle\alpha|-\alpha\rangle|\ll1$ is satisfied.

Next, we note that the direct product of two cat states can be expressed as the sum of two entangled cat states with different parities,
\begin{eqnarray}
|\psi\rangle&=&\frac{1}{2+\epsilon_{\rm sc}}\left[(|\alpha\rangle+|-\alpha\rangle)\otimes(|\alpha\rangle+|-\alpha\rangle)\right]\nonumber\\
            &\equiv&\frac{1}{\sqrt{2+\epsilon_{\rm eo}}}[|{\rm even}\rangle+|{\rm odd}\rangle],\label{directproductcat}
\end{eqnarray}
where the entangled cat states with even and odd parity, respectively, read
\begin{align}
|{\rm even}\rangle & \equiv \frac{1}{\sqrt{2+\epsilon_{\rm ec}}}\left(|\alpha\rangle\otimes|\alpha\rangle+|-\alpha\rangle\otimes|-\alpha\rangle\right) ,\tag{8a} \label{Eq:even_parity_ecs}\\
|{\rm odd}\rangle & \equiv \frac{1}{\sqrt{2+\epsilon_{\rm ec}}}\left(|\alpha\rangle\otimes|-\alpha\rangle+|-\alpha\rangle\otimes|\alpha\rangle\right) .\tag{8b}\label{Eq:odd_parity_ecs}
\end{align}
The normalization factor of the total state in Eq.~(\ref{directproductcat}) is modified by a small quantity $\epsilon_{\rm eo}$, as the entangled cat states with different parities are not orthogonal to each other. For a similar reason, the correction $\epsilon_{\rm ec}$ is necessary in Eqs.~(\ref{Eq:even_parity_ecs}) and (\ref{Eq:odd_parity_ecs}). The even parity state $|{\rm even}\rangle$ is a dark state of the dissipator~(\ref{couplingH}), while the odd one is not. Therefore, the steady state of the system is the entangled cat state with even parity under the influence of $H$, $\mathcal{L}_{\rm d}$, and $\mathcal{L}_{\rm c}$.

\subsection{Influence of single-photon loss}

The single-photon dissipation generically has a negative effect on the generation of entangled cat states. It is straightforward to check the relations
\begin{eqnarray}
\mathcal{L}_{\rm s}(|{\rm even}\rangle\langle{\rm even}|)&\neq&0,\nonumber\\
\mathcal{L}_{\rm s}(|{\rm odd}\rangle\langle{\rm odd}|)&\neq&0,
\end{eqnarray}
which imply that the steady state is, when $\mathcal{L}_{\rm s}$ is included, not an entangled cat state. In addition, both the pure state $|\rm even\rangle$ and the mixed state
\begin{eqnarray}\label{classicalmix}
|\alpha\rangle\langle\alpha|\otimes|\alpha\rangle\langle\alpha|+|-\alpha\rangle\langle-\alpha|\otimes|-\alpha\rangle\langle-\alpha|
\end{eqnarray}are steady states of the system in absence of the single-photon dissipation. Therefore, the effect of a sufficiently weak single-photon loss $\mathcal{L}_{\rm s}$ is to destroy the coherence. For a strong single-photon loss, the steady state becomes a squeezed vacuum state~\cite{CIM5}. This parameter regime, which is below threshold, is not considered here. In the regime of medium single-photon loss, i.e., slightly above threshold, the dissipative coupling can pick up the wrong parity~\cite{CIMstability}, so that the qualities of entangled cat states can be significantly reduced. However, it is still possible to obtain a cat-like transient state if the single-photon dissipation is within a proper range.

\section{Entanglement detection}\label{criterion}

To certify the presence of entanglement in the generated states, we apply an entanglement criterion which is formulated in terms of modular variables~\cite{mec1,mec2,mec3,mec4,mec5}. By virtue of these, the criterion is sensitive to periodic structures in the states, which comprises entangled cat states as the two-component case. On the other hand, the criterion is not sensitive to the ``internal'' structure of the repeating state component, e.g., if it is Gaussian or not. This is relevant here, because we can only assume coherent-state components in the ideal case.

We now briefly recapitulate the steps towards the evaluation of the cat state-sensitive entanglement criterion. The criterion is based on the measured joint position and momentum distributions,
\begin{eqnarray} \label{Eq:position_and_momentum_distribution}
P_x(x_1,x_2)&=&\langle x_1|\langle x_2|\rho|x_1\rangle|x_2\rangle,\nonumber\\
P_p(p_1,p_2)&=&\langle p_1|\langle p_2|\rho|p_1\rangle|p_2\rangle,\label{xpdistribution}
\end{eqnarray}
with $|x_k\rangle$ and $|p_k\rangle$ the ``position'' and ``momentum'' (i.e., conjugate quadratures) eigenstates of the two modes, respectively. For simplicity, we assume that the frequency $\omega$ and the effective mass $m$ satisfy $\omega m=1$ for the DOPO modes, so that we define the position and momentum operators as
\begin{eqnarray} \label{Eq:position_momentum_operator}
\hat{x}_k&=&\frac{1}{\sqrt{2}}(a_k+a_k^{\dag}),\nonumber\\
\hat{p}_k&=&\frac{1}{\sqrt{2}i}(a_k-a_k^{\dag}),
\end{eqnarray}
where $k=1,2$. By introducing a length scale $\l_x$ and an associated momentum scale $\l_p$ with $l_xl_p=2\pi$, we now redefine the position and momentum eigenvalues in terms of integer and modular parts,
\begin{eqnarray} \label{Eq:modular_variables}
x_k&=&N_{x,k}l_x+\bar{x}_k , \nonumber \\
p_k&=&N_{p,k}l_p+\bar{p}_k ,
\end{eqnarray}
with the integer components $N_{x,k}$ and $N_{p,k}$, and the modular rest components $\bar{x}_k\in[0,l_x)$ and $\bar{p}_k\in [0,l_p)$. While the length scale can, in principle, be chosen arbitrarily, there exists an optimal choice of $l_{x}$ (or $l_p$), which depends on the phase-space separation between different state components. Note that the relation $l_xl_p=2\pi$ reflects the relation between the separation of the state components (be it in position or momentum space) and the periodicity of the associated interference pattern in the conjugate variable.

To assess if the measurement data (\ref{Eq:position_and_momentum_distribution}) implies entanglement, we now determine, from this data, the distributions of collective variables that are derived from the decompositions (\ref{Eq:modular_variables}), e.g., the distribution of the total modular position $(\bar{x}_1+\bar{x}_2)$. Such postprocessing of the measurement data is always possible. The optimal choice of the two required collective variables depends on the form of the state that underlies the measurement data (\ref{Eq:position_and_momentum_distribution}). In our case, where, due to pure imaginary amplitudes $\alpha$ (in the ideal case), the macroscopic superposition is laid out in the momentum coordinate, the appropriate collective variables are the total modular position and the relative integer momentum,
\begin{eqnarray} \label{Eq:collective_modular_variables}
\bar{x}_{\rm tot}&=&\bar{x}_{1}+\bar{x}_{2},\nonumber\\
N_{p,\rm rel}&=&{N}_{p,1}-{N}_{p,2}.
\end{eqnarray}
The distributions of $\bar{x}_{\rm tot}$ and $N_{{p,{\rm rel}}}$ can be calculated with the distributions (\ref{xpdistribution}),
\begin{eqnarray}
P_{\bar{x}_{\rm tot}}&=&\sum_{N_{x,1},N_{x,2}}\int_0^{l_x} dxP_x(N_{x,1}l_x+\bar{x}_{\rm tot}-x,N_{x,2}l_x+x),\nonumber\\
P_{N_{p,{\rm rel}}}&=&\sum_{N}\int_0^{l_p} dp_1dp_2P_p[(N_{p,{\rm rel}}+N)l_p+p_{1},Nl_p+p_2],\nonumber\\
\end{eqnarray}
The entanglement criterion is now formulated in terms of the variances that follow from the distributions of the collective variables (\ref{Eq:collective_modular_variables}). Specifically, the {\it modular entanglement criterion} (mec)~\cite{mec2} states that the state underlying the measurement data (\ref{Eq:position_and_momentum_distribution}) must be entangled if
\begin{eqnarray} \label{Eq:modular_entanglement_criterion}
C_{\rm mec}\equiv\langle(\Delta N_{\rm p,rel})^2\rangle+\langle(\Delta \bar{x}_{\rm tot}/l_x)^2\rangle\leq C_{\rm et},
\end{eqnarray}
i.e., if the sum of variances $C_{\rm mec}$ remains below the entanglement threshold value $C_{\rm et} \approx 0.1565$ \cite{mec1}. The latter follows from a state-independent additive uncertainty relation for the modular variables, $\langle{(\Delta N_p)^2_j}\rangle+\langle{(\Delta \bar{x})^2_j}\rangle\geq C_{\rm et}/2$. The criterion (\ref{Eq:modular_entanglement_criterion}), which is a sufficient condition for entanglement, is similar to a variance-based entanglement criterion in terms of standard (i.e., not modular) variables, which is sensitive to the entanglement in bipartite Gaussian states~\cite{gse1,gse2}. Note that a suppressed variance $(\Delta \bar{x}_{\rm tot})^2$ of the total modular position reflects the presence of a fringe pattern in the distribution of the total position $x_{\rm tot}=x_1+x_2$. While we cannot exclude that the state is entangled if the criterion is not satisfied (e.g., entangled Gaussian states never satisfy the criterion), we can expect that any cat state-like entanglement is reliably detected. For instance, the criterion (\ref{Eq:modular_entanglement_criterion}) may also be used to verify the continuous-variable entanglement in alternative cat-state generation schemes~\cite{expecs3}.
\subsection{Modular entanglement criterion examples}
\subsubsection{Even-parity entangled cat state}
Let us consider three instructive examples to see how the modular entanglement criterion works. First, let us assume that the state underlying the measured data (\ref{Eq:position_and_momentum_distribution}) is our ideal target state $\frac{1}{N_{\rm e}}(|\alpha\rangle\otimes|\alpha\rangle+|-\alpha\rangle\otimes|-\alpha\rangle)$, with an imaginary (i.e., the macroscopic superposition is laid out in momentum) amplitude $\alpha$, [cf.~(\ref{Eq:even_parity_ecs})]. To detect the entanglement in this state, we use the total modular position variable and the relative integer momentum variable, cf.~(\ref{Eq:collective_modular_variables}). The optimal choice for the modular length scale $l_p$ is $\rm{Re}(2\sqrt{2}\alpha)$ [the factor $\sqrt{2}$ can be traced back to the transformation to position and momentum operators, cf.~(\ref{Eq:position_momentum_operator})]. For sufficiently large $|\alpha|$, the variance of the relative integer momentum is negligibly small, $(\Delta N_{p,{\rm rel}})^2 \approx 0$, while the variance of the total modular position, due to the presence of an interference pattern, becomes $(\Delta \bar{x}_{\rm tot}/l_x)^2 \approx 0.1167$ \cite{mec1}. We thus obtain for the sum of variances $C_{\rm mec} \approx 0.1167 < C_{\rm et} \approx 0.1565$, and the state is certified as entangled by the modular entanglement criterion (\ref{Eq:modular_entanglement_criterion}). Note that the value $C_{\rm mec} \approx 0.1167$ is the smallest possible that can be achieved with two-component macroscopic superpositions; therefore, in our realistic setting, $C_{\rm mec}$ will always exceed this value, due to single-photon dissipation and finite $|\alpha|$.
\subsubsection{Odd-parity entangled cat state}
As the second example, let us consider the odd-parity entangled cat state $\frac{1}{N_{\rm o}}(|\alpha\rangle\otimes|-\alpha\rangle+|-\alpha\rangle\otimes|\alpha\rangle)$, [cf.~(\ref{Eq:odd_parity_ecs})]. The appropriate variables are now the total integer momentum $N_{p,\rm tot}={N}_{p,1}+{N}_{p,2}$ and the relative modular position $\bar{x}_{\rm rel}=\bar{x}_{1}-\bar{x}_{2}$, while the optimal choice for the length scale remains $l_p = \rm{Re}(2\sqrt{2}\alpha)$. Under these conditions, the modular entanglement criterion again evaluates as $C_{\rm mec} \approx 0.1167$ for $|\alpha| \gg 1$, and the state's entanglement is certified.
\subsubsection{Separable classically correlated mixed state}
Thirdly, we consider the separable, but classically correlated, mixed state in Eq.~(\ref{classicalmix}). We can then again choose the variables (\ref{Eq:collective_modular_variables}) to obtain a strongly suppressed variance of the relative integer position, $(\Delta N_{\rm p,rel})^2 \approx 0$ for $|\alpha| \gg 1$. However, due to the absence of a fringe pattern in the distribution of $x_{\rm tot}=x_1+x_2$, the variance of the total modular position becomes $(\Delta \bar{x}_{\rm tot}/l_x)^2 \approx 0.167$ \cite{mec1}, and the modular entanglement criterion (\ref{Eq:modular_entanglement_criterion}) does, correctly, not detect entanglement.

\subsection{Modular variables during evolution}

Assigning to a set of model parameters a unique and optimal entanglement qualifier $C_{\rm mec}$ requires a time-adaptive evaluation of $C_{\rm mec}$, which we detail in the following. The initial state is assumed to be the vacuum state $|0\rangle \otimes |0\rangle$, and the time evolution is governed by the Hamiltonian~(\ref{systemhamiltonian}) and all three dissipative contributions summarized in (\ref{Eq:dissipative_contributions}). The two-photon dissipation rate $\gamma_{\rm d}$ is set to be $1$, and all the other parameters are provided as the ratio to $\gamma_{\rm d}$. While the optimal length scale for the ideal asymptotic state is $l_{p} = 4\sqrt{S/\gamma_{\rm d}}$, it can, under realistic conditions, take other values. Therefore, we determine the optimal $l_p$ for every choice of model parameters and at every point in time.
\begin{figure}[t]
\center
\includegraphics[width=3.5in]{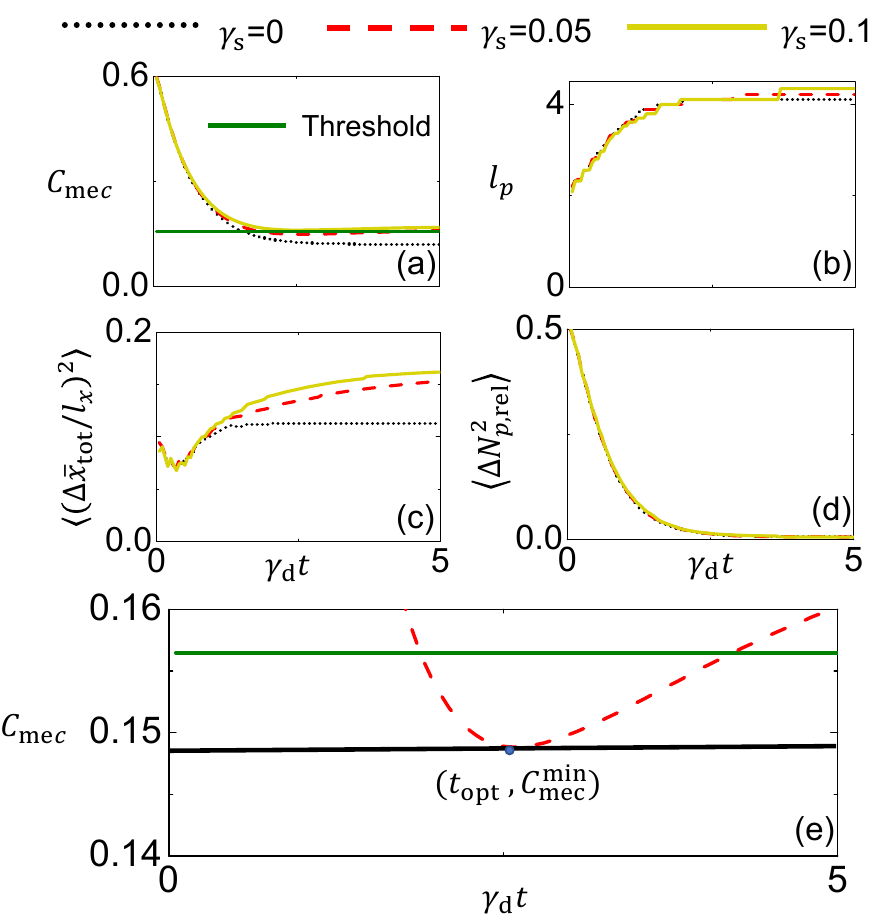}
\caption{Time-adapted evaluation of the modular entanglement criterion. The pump intensity $S$ and the two-photon dissipation rate $\gamma_{\rm d}$ are set to be $1$. All the other parameters are expressed as the ratios to $\gamma_{\rm d}$. The collective dissipation rate $\gamma_{\rm c}$ is set to be $10\gamma_{\rm d}$. (a) Instantaneous sum of variances $C_{\rm mec}$. (b) Time-adapted optimal momentum scale $l_p$. We set the optical frequency $\omega=1$ to better show the relation between $l_p$ and cat state separation $\alpha$. (c) Instantaneous variance of the total modular position $\bar{x}_{\rm tot}$. (d) Instantaneous variance of the relative integer momentum $N_{p,\rm rel}$. (e) For a given set of model parameters, the minimizing instantaneous entanglement qualifier $C_{\rm mec}$ is taken to assess the detectable entanglement for this model parameter set. This is shown for the example of $\gamma_s=0.05$.}
\label{fig1}
\end{figure}

In Figure~\ref{fig1} we demonstrate how the mechanism underlying the state generation unfolds in the temporal evolution of the variances of the collective variables (\ref{Eq:collective_modular_variables}). The small jumps of the length scale $l_p$ in Figure~\ref{fig1}(b) are due to the discrete set of scanned values. Figure~\ref{fig1}(a) shows the evolution of the entanglement qualifier $C_{\rm mec}$ in relation to the entanglement detection threshold for different choices of the single-photon dissipation rate $\gamma_s$. In absence of single-photon dissipation $\gamma_{\rm s}$ (black dotted curve), the steady state is the entangled cat state (\ref{Eq:even_parity_ecs}). However, single-photon dissipation significantly modifies the evolution. If this dissipative process is sufficiently weak (red dashed curve), the entanglement qualifier $C_{\rm mec}$ still temporarily drops below the entanglement detection threshold. With increasing $\gamma_{\rm s}$, however, the minimum $C_{\rm mec}$ eventually exceeds the threshold, and entanglement cannot be detected at any time.

Additional insights can be drawn from the unfoldings of $l_p$, $\langle(\Delta \bar{x}_{\rm tot}/l_x)^2\rangle$, and $\langle(\Delta N_{p,{\rm rel}})^2\rangle$, as depicted in Figs.~\ref{fig1} (b)-(d). The increasing $l_p$ in Fig.~\ref{fig1}(b) informs us about the growing amplitudes towards their steady-state values. In absence of single-photon dissipation, we find the steady-state value of the ideal entangled cat state, $l_p = 4\sqrt{S/\gamma_{\rm d}}$. For other cases, we find slight deviations from this value.

The temporal course of the variance of the total modular position $\bar{x}_{\rm tot}$ is shown in Fig.~\ref{fig1}(c). This variance is related to the interference pattern, and hence is highly sensitive to the single-photon dissipation. In absence of single-photon loss, this variance assumes an asymptotic value below the entanglement detection threshold (black dotted curve); in the other cases, it grows, after taking a minimum, towards the ``no interference" value 0.167 (red dashed curve and yellow solid curve). We can interpret this variance as an indicator for the coherence in the state. Figure~\ref{fig1}(d) shows the course in time of the variance of the relative integer momentum $\bar{N}_{p,\rm rel}$, which captures the correlations between the macroscopic state components of the two modes. Since these correlations are also present in the decohered state (cf.~the discussion of our third example above), this variance is not significantly affected by the single-photon dissipation rate. After the analysis of $\langle(\Delta \bar{x}_{\rm tot}/l_x)^2\rangle$ and $\langle(\Delta N_{\rm p,rel})^2\rangle$, we come to a counterintuitive result. In Fig.~\ref{fig1}(b), the optimal $l_p$ increases with increasing single-photon dissipation. This is because larger $l_p$ can decrease the variance of $N_p$, and thus can become favorable in the presence of decoherence, since the latter neutralizes the detrimental effect of the length-scale offset on the variance of the interference-sensitive variance of $\overline{x}$.

As we have shown, in the presence of single-photon dissipation, $C_{\rm mec}$ assumes a minimum before it monotonically grows towards its asymptotic value. We thus can take this minimum value of $C_{\rm mec}$ as a qualifier for detectable entanglement. This is exemplified in Fig.~\ref{fig1}(e). From now on, we use this minimum value $C_{\rm mec}^{\rm min}$ to assess the potential to achieve cat-state entanglement for different model parameter choices. Note that $C_{\rm mec}^{\rm min}$ does not qualify as an entanglement measure, and thus lower values of $C_{\rm mec}^{\rm min}$ (below the threshold) do not necessarily indicate higher entanglement.

\section{Conditions for entanglement generation and detection}\label{parareg}

In the previous section, we demonstrated the possibility to generate detectable entangled cat states in DOPOs, and detailed the underlying state formation. The preliminary insights from Fig.~\ref{fig1} indicate that the generation of entangled cat states is challenging in DOPOs, due to the requirement of a small rate ratio $\gamma_{\rm s}/\gamma_{\rm d}$. In the following, we perform a systematic analysis in parameter space, to determine the {\it minimum requirements for the successful generation of entangled cat states}. This is achieved by finding the minimum $C_{\rm mec}$ for each set of parameters.
\begin{figure}[t]
\center
\includegraphics[width=3.2in]{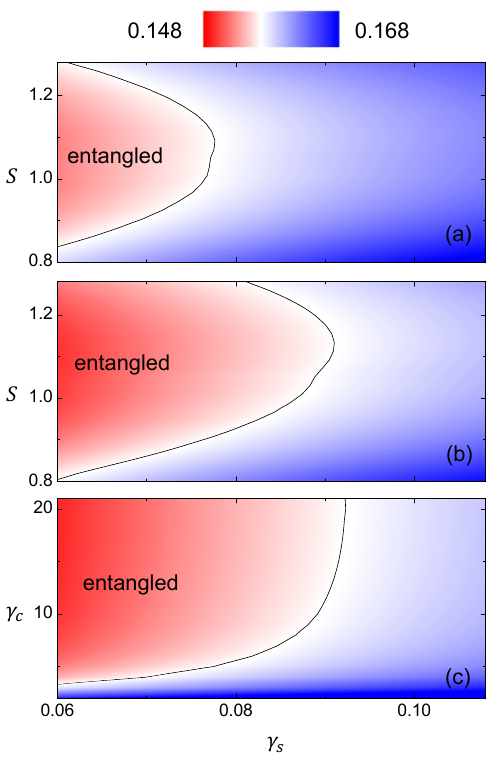}
\caption{Cat-state entanglement qualifier $C^{\rm min}_{\rm mec}$ for different model parameters. Here, $S$ is the pump intensity, $\gamma_s$ is single-photon loss rate, and $\gamma_c$ is collective dissipation rate. Parameter combinations that give rise to a $C^{\rm min}_{\rm mec}$ below the the entanglement detection threshold, i.e., detectable cat-state entanglement, are shown in red; while the parameter combinations that give rise to a $C^{\rm min}_{\rm mec}$ above threshold are shown in blue. The black solid curves demarcate the approximate boundary between the cat state-entanglement certifiable parameter choices and the inconclusive parameter choices. The two-photon dissipation intensity $\gamma_{\rm d}$ is set to be $1$. All other parameters are expressed as ratios to $\gamma_{\rm d}$. The collective dissipation rates are $\gamma_{\rm c}=5\gamma_{\rm d}$ and $\gamma_{\rm c}=10\gamma_{\rm d}$ for (a) and (b), respectively. (c) shows the results for different values of $\gamma_{\rm c}$, with $S=1.05\gamma_{\rm d}$.}
\label{fig2}
\end{figure}

\begin{figure}[t]
\center
\includegraphics[width=3.5
in]{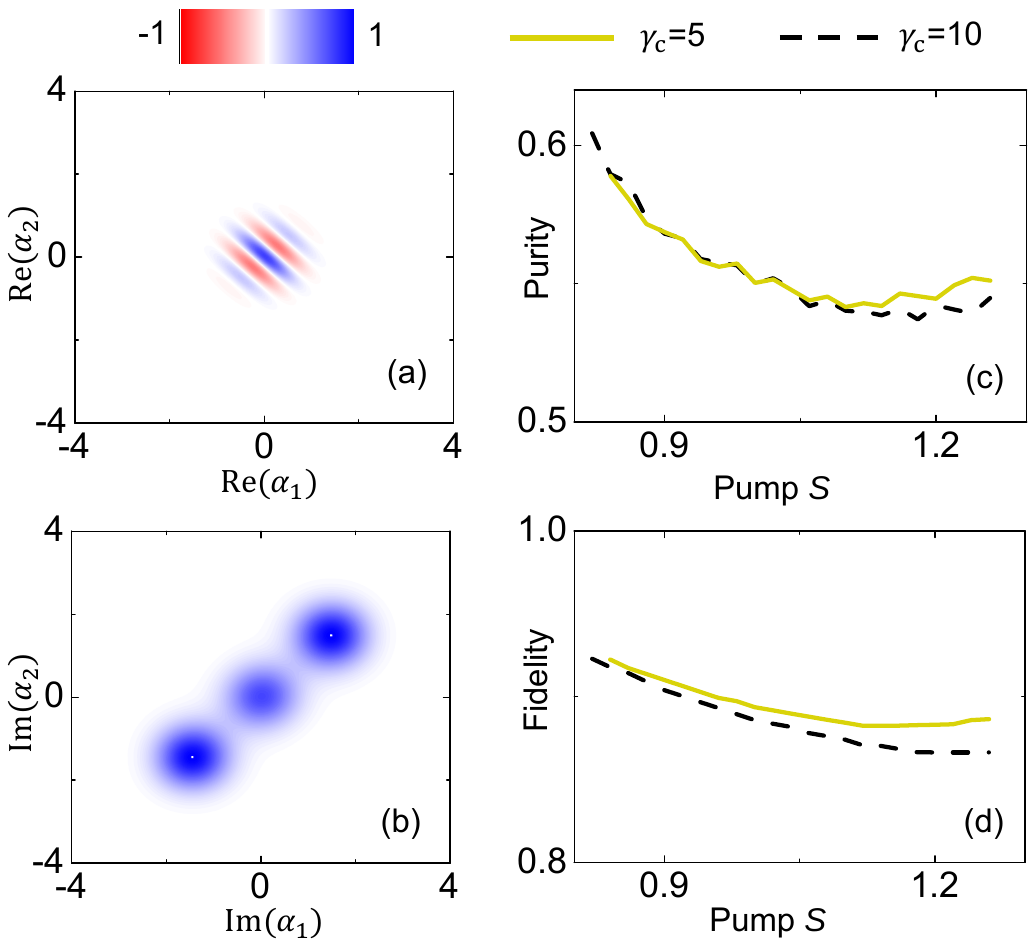}
\caption{Properties of the states near the boundary. (a) The section of the joint Wigner function on the real plane. (b) The section of the joint Wigner function on the imaginary plane. (c) Purity of the state, with minimum $C_{\rm mec}$ during the evolution, for different pump intensities $S$ near the boundary. (d) Highest cat state fidelity, which can be achieved in single DOPO with parameters $S$, $\gamma_{\rm d}$, and $\gamma_{\rm s}$ near the boundary.}
\label{fig3}
\end{figure}

In Figs.~\ref{fig2}(a) and \ref{fig2}(b) we analyze the interplay between the pump intensity $S$ and the single-photon dissipation $\gamma_{\rm s}$ with respect to the entanglement generation. To reach below the threshold under the influence of dissipation, there is an optimal choice of the pump intensity $S$ in both figures. Such a value is decided by the trade off between coherence and the separation, which are both key properties of a high quality entangled cat state. When the pump is strong, the average photon number is large. Two components of such a ``large" state have less overlap [smaller $\langle(\Delta N_{\rm p,rel})^2\rangle$], but the system can suffer stronger single-photon-loss effects. As we have shown in Fig.~\ref{fig1}(c), this loss can increase $\langle(\Delta \bar{x}_{\rm tot}/l_x)^2\rangle$. With a small average photon number, the $\langle(\Delta \bar{x}_{\rm tot}/l_x)^2\rangle$ can be small due to lower probability to loss photons. However, the large overlap can result in a large $\langle(\Delta N_{\rm p,rel})^2\rangle$. The optimal value of $S$ slightly changes with $\gamma_c$, but is around $1.05\gamma_{\rm d}$ for both the $\gamma_c=10$ case and the $\gamma_c=5$ case. It is not surprising to see a larger area with entanglement in Fig.~\ref{fig2}(b) compared to Fig.~\ref{fig2}(a), because the collective dissipation rate $\gamma_{\rm c}$ is stronger. As $\gamma_{\rm c}$ can be conceptually very strong, we further check how much it can improve the results in Fig.~\ref{fig2}(c). It is confirmed in Fig.~\ref{fig2}(c) that increasing $\gamma_{\rm c}$ always has positive effects. However, such effects become insignificant if $\gamma_{\rm c}$ is larger than $10\gamma_{\rm d}$.

Although the modular entanglement criterion provides a concise description of entanglement, the information of the state is incomplete. Next, we consider the state with entanglement near the boundary in Figs.~\ref{fig2}(a) and (b). The sections of joint Wigner function, purity of the state, and the corresponding cat state in a single DOPO are shown in Fig.~\ref{fig3}. The joint Wigner function is a tool to provide some intuitive pictures of the entangled cat states~\cite{expecs3} by calculating the distribution on the coherent state basis $|\alpha_1\rangle\otimes|\alpha_2\rangle$ of the two entangled modes. Such a function is similar to a Wigner function~\cite{Wignerfunction1} but in a four-dimensional space, so that only important sections instead of the full function are plotted. In Fig.~\ref{fig3}(a), the section on the real plane $\rm{Im}(\alpha_1)=\rm{Im}(\alpha_2)=0$, which refers to the interference pattern, is shown. This pattern can reveal the coherent superposition of several components, but does not assure entanglement. Figure~\ref{fig3}(b) provides the section on the imaginary plane $\rm{Re}(\alpha_1)=\rm{Re}(\alpha_2)=0$. This figure is very similar to the Wigner function of a cat state, and clearly shows the two components of the entangled cat state. The top right disk corresponds to the $|\alpha\rangle|\alpha\rangle$ component, and the bottom left one corresponds to the $|-\alpha\rangle|-\alpha\rangle$ component. Note that the disk in the center is a section of the interference pattern instead of an additional component. This central disk can become negative for an entangled cat state with odd parity~\cite{expecs3}. Figure~\ref{fig3}(c) shows the purity of the state with minimum $C_{\rm mec}$ during the evolution. Oscillations in Fig.~\ref{fig3}(c) are the result of choosing an optimal modular length, which has limited influence on the main trends. The purity changes with pump intensity $S$ but the variation is only about $0.05$. Generally speaking, purity decreases with increasing $S$, because larger separation brings more tolerance for poor coherence. However, purity can increase with pump intensity for large $S$ when the collective dissipation rate is weak (yellow solid curve). The pump term creates the entangled cat states with both parities. A weak $\gamma_{\rm c}$ can result in a larger probability of the state with unwanted parity. This state contributes to the purity, but has negative effects on entanglement.

With the results in Figs.~\ref{fig3}(a)-(c), some basic information of the entangled state near threshold is provided. Then, we try to compare it to the single-mode case, which is well discussed in other works~\cite{dpocat1,dpocat2,dpocat3,dpocat4,dpocat5}. The highest fidelity of the cat state, which can be obtained with the single-mode parameters in Fig.~\ref{fig3}(c), is shown in Fig.~\ref{fig3}(d). It is intuitive to expect a similar parameter dependence in the single-mode case and the entangled case, because the main detrimental effect in both cases is the single-photon dissipation. The results in Fig.~\ref{fig3}(d) also agrees with this expectation, which shows the same trend as the curves in Fig.~\ref{fig3}(c). Without the modular length choice, there is no jump in Fig.~\ref{fig3}(d). We also find that the cat state fidelity with boundary parameters is below $0.9$ for most values of $S$. Therefore, the entangled cat state can be generated if we can access cat states with fidelity higher than $0.9$ in single DOPO.
\begin{figure}[t]
\center
\includegraphics[width=2.8
in]{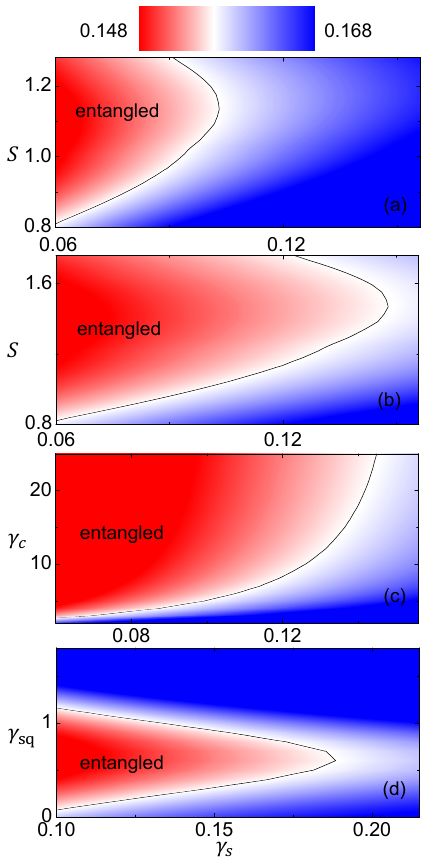}
\caption{The value of $C^{\rm min}_{\rm mec}$ for a squeezed environment. The area below threshold is marked as red, and the one above is marked as blue. The black solid curve is the approximate boundary between the entangled area and unentangled area. The two photon dissipation intensity $\gamma_{\rm d}$ is set to be $1$. All the other parameters are expressed as the ratio to $\gamma_{\rm d}$. (a) The influence of $S$ and $\gamma_{\rm s}$ with $\gamma_{\rm c}=5\gamma_{\rm d}$ and $\gamma_{\rm sq}=1$. (b) The influence of $S$ and $\gamma_{\rm s}$ with $\gamma_{\rm c}=10\gamma_{\rm d}$ and $\gamma_{\rm sq}=1$. (c) The influence of $\gamma_{\rm c}$ and $\gamma_{\rm s}$ with $\gamma_{\rm sq}=1$ and $S=1.05\gamma_{\rm d}$. (d) The influence of $\gamma_{\rm sq}$ and $\gamma_{\rm s}$ with $\gamma_{\rm c}=10\gamma_{\rm d}$ and $S=1.2\gamma_{\rm d}$.}
\label{fig4}
\end{figure}
\section{Environmental engineering}\label{bathsq}
In the previous section, we analyzed the conditions for generating entangled cat states in coupled DOPOs. It is obvious that these parameters are very challenging. The cat state in DOPO, which is less difficult, is also hard to access for now. Therefore, control methods might be necessary for the creation of such quantum states in DOPOs. One potential choice is environmental engineering~\cite{CIM9,sqb1,disspativestabilization,sqb2}, which can significantly reduce the influence of single-photon dissipation. Below, we show that this approach for a single-mode system can also be applied to generate entangled states. Squeezing is only introduced in the single-photon dissipation channel, while other dissipation terms are unaffected. The squeezed single-photon dissipation term has the following form,
\begin{eqnarray}
\mathcal{L}^{\rm sq}_{\rm s}(\rho)&=&\left(1+N\right)\sum_{k=1}^2\mathcal{L}(\gamma_s,a_k,\rho)+N\sum_{k=1}^2\mathcal{L}(\gamma_s,a_k^{\dag},\rho)\nonumber\\
                                 &&-M\sum_{k=1}^2\frac{\gamma_{\rm s}}{2}\left(2a_k\rho a_k-\{a_ka_k,\rho\}\right)\nonumber\\
                                 &&-M^*\sum_{k=1}^2\frac{\gamma_{\rm s}}{2}\left(2a_k^{\dag}\rho a_k^{\dag}-\{a_k^{\dag}a_k^{\dag},\rho\}\right).
\end{eqnarray}
Here, the two parameters are $N={\rm sinh}^2(\gamma_{\rm sq})$ and $M={\rm sinh}(\gamma_{\rm sq}){\rm cosh}(\gamma_{\rm sq})e^{-i\theta_{\rm sq}}$, with the intensity of squeezing $\gamma_{\rm sq}$ and the phase $\theta_{\rm sq}$. The phase term should fit the orientation of the entangled cat state, which is set to be $\pi$ in our case.

In Fig.~\ref{fig4}, we show the results with a squeezed environment. Figures~\ref{fig4}(a) and (b) correspond to the situations in Figs.~\ref{fig2} (a) and (b), respectively. We find that the regime for detectable entanglement is enlarged for both the strong collective dissipation case and the weak collective dissipation case. Another difference is the optimal value for the pump intensity $S$, which increases under the influence of the squeezed dissipation channel. Squeezing results in weaker $x$ fluctuations at the cost of stronger $p$ fluctuations. Therefore, a larger $S$ is necessary, so that a larger $p$ separation can compensate for the fluctuation. Figure~\ref{fig4}(c) shows the influence of $\gamma_{\rm c}$ and $\gamma_{\rm s}$. In Fig.~\ref{fig2}(c) the effect of $\gamma_{\rm c}$ saturates around $\gamma_{\rm c}=10\gamma_{\rm d}$, but increasing $\gamma_{\rm c}$ can improve the performance after passing $20\gamma_{\rm d}$ in the squeezed case. We study the influence of the squeezing parameter $\gamma_{\rm sq}$ in Fig.~\ref{fig4}(d). There is an optimal value of $\gamma_{\rm sq}$, which is about $0.5$ for the parameters in Fig.~\ref{fig4}(d).

The compatibility of the dissipative coupling with the single-mode environment engineering is shown with the results in Fig.~\ref{fig4}. Although the parameters are not the optimal ones, the improvement obtained is significant. If high quality cat states can be realized with control methods, then these approaches are very likely also effective for the production of the entangled cat states.

\section{conclusions}\label{conclusions}

We studied the generation of entangled cat states in DOPOs with collective dissipation. The quality of the state is characterized by the modular entanglement criterion, which can detect the entanglement in cat-like states. Based on this criterion, we identify the necessary parameter regimes to observe entangled cat states. Our results also clarify the influence of different parameters and the optimal ones to access the desired entangled state. To better relate the problem considered to the existing works, we compared the threshold cases with the single-mode cat state situations. Although the choice of optimal parameters can be quite complicated, the fidelity of the corresponding single-mode state does not change much. As a result, creating an entangled cat state is comparable to the problem of generating a cat state with sufficient quality, if the collective dissipation method is applied. Due to the challenging parameters in DOPOs, quantum control may be necessary. Therefore, we also considered environment engineering as an example. According to our numerical results, the entanglement creating method can cooperate well with the single-mode control method. Our work presents the conditions for achieving entangled cat states in DOPOs, and relates this task to its single-mode counterpart. In this sense, it may, e.g., help to extend the application of the coherent Ising machine into the quantum regime.

\section{Acknowledgments}
J.Q.Y. is partially supported by the National Key Research and Development Program of China (Grant No. 2016YFA0301200) and the National Natural Science Foundation of China (NSFC) (Grant No. 11774022 and No. U1801661). F.N. is supported in part by: Nippon Telegraph and Telephone Corporation (NTT) Research, the Japan Science and Technology Agency (JST) [via the Quantum Leap Flagship Program (Q-LEAP), the Moonshot R\&D Grant Number JPMJMS2061, and the Centers of Research Excellence in Science and Technology (CREST) Grant No. JPMJCR1676], the Japan Society for the Promotion of Science (JSPS) [via the Grants-in-Aid for Scientific Research (KAKENHI) Grant No. JP20H00134 and the JSPS-CRFBR Grant No. JPJSBP120194828], the Army Research Office (ARO) (Grant No. W911NF-18-1-0358), the Asian Office of Aerospace Research and Development (AOARD) (via Grant No. FA2386-20-1-4069), and the Foundational Questions Institute Fund (FQXi) via Grant No. FQXi-IAF19-06.


%


\begin{references}
%
%
\bibitem{introcat1}S. Haroche, J.-M Raimond, Exploring the Quantum: Atoms, Cavities, And Photons, Oxford Univ. Press (2006).
%
%
\bibitem{introcat2}T. Ogawa, M. Ueda, and N. Imoto, Generation of the Schr\"odinger-cat state by continuous photodetection, \href{https://doi.org/10.1103/PhysRevA.43.6458}{Phys. Rev. A {\bf 43}, 6458(R) (1991)}.
%
%
\bibitem{introcat3}M. Brune, E. Hagley, J. Dreyer, X. Ma$\rm{\hat{\i}}$tre, A. Maali, C. Wunderlich, J. M. Raimond, and S. Haroche, Observing the Progressive Decoherence of the ``Meter'' in a Quantum Measurement, \href{https://doi.org/10.1103/PhysRevLett.77.4887}{Phys. Rev. Lett. {\bf 77}, 4887 (1996)}.
%
%
\bibitem{introcat4} Y.-x. Liu, L. F. Wei, and F. Nori, Preparation of macroscopic quantum superposition states of a cavity field via coupling to a superconducting charge qubit, \href{https://doi.org/10.1103/PhysRevA.71.063820}{Phys. Rev. A {\bf71}, 063820 (2005)}.
%
%
\bibitem{introcat5}S. Del$\rm\acute{e}$glise, I. Dotsenko, C. Sayrin, J. Bernu, M. Brune, J.-M. Raimond, and S. Haroche, Reconstruction of non-classical cavity field states
with snapshots of their decoherence, \href{https://doi.org/10.1038/nature07288}{Nature {\bf 455}, 510 (2008)}.
%
%
\bibitem{introcat6}M. Kira, S. W. Koch, R. P. Smith, A. E. Hunter, and S. T. Cundiff, Quantum spectroscopy with Schr$\rm\ddot{o}$dinger-cat states, \href{https://doi.org/10.1038/NPHYS2091}{Nat. Phys. {\bf7}, 799 (2011)}.
%
%
\bibitem{ecs1}B. C. Sanders, Entangled coherent states, \href{https://doi.org/10.1103/PhysRevA.45.6811}{Phys. Rev. A {\bf 45}, 6811 (1992)}.
%
%
\bibitem{ecs2}F. D. Martini, M. Fortunato, P. Tombesi, and D. Vitali, Generating entangled superpositions of macroscopically distinguishable states within a parametric oscillator, \href{https://doi.org/10.1103/PhysRevA.60.1636}{Phys. Rev. A {\bf 60}, 1636 (1999)}.
%
%
\bibitem{ecs3}M. Paternostro, M. S. Kim, and B. S. Ham, Generation of entangled coherent states via cross-phase-modulation in a double
electromagnetically induced transparency regime, \href{https://doi.org/10.1103/PhysRevA.67.023811}{Phys. Rev. A {\bf 67}, 023811 (2003)}.
%
%
\bibitem{ecs4}H. Jeong and N. B. An, Greenberger-Horne-Zeilinger-type and W-type entangled coherent states:
Generation and Bell-type inequality tests without photon counting, \href{https://doi.org/10.1103/PhysRevA.74.022104}{Phys. Rev. A {\bf 74}, 022104 (2006)}.
%
%
\bibitem{ecs5}C. Arenz, C. Cormick, D. Vitali, and G, Morigi, Generation of two-mode entangled states
by quantum reservoir engineering, \href{https://doi.org/10.1088/0953-4075/46/22/224001}{J. Phys. B: At. Mol. Opt. Phys. {\bf 46}, 224001 (2013)}.
%
%
\bibitem{ecs6}B. Xiong, X. Li, S.-L. Chao, Z. Yang, W.-Z. Zhang, and L. Zhou, Generation of entangled Schr${\rm \ddot{o}}$dinger cat state
of two macroscopic mirrors, \href{https://doi.org/10.1364/OE.27.013547}{Opt. Express {\bf 27}, 13547 (2019)}.
%
%
\bibitem{expecs1}W.-B. Gao, C.-Y. Lu, X.-C. Yao, P. Xu, O. G${\rm \ddot{u}}$hne, A. Goebel, Y.-A. Chen, C.-Z. Peng, Z.-B. Chen, and J.-W. Pan, Experimental demonstration of a hyper-entangled
ten-qubit Schr\"odinger cat state, \href{https://doi.org/10.1038/NPHYS1603}{Nat. Phys. {\bf 6}, 331 (2010)}.
%
%
\bibitem{expecs2}X.-C. Yao, T.-X. Wang, P. Xu, H. Lu, G.-S. Pan, X.-H. Bao, C.-Z. Peng,
C.-Y. Lu, Y.-A. Chen, and J.-W. Pan, Observation of eight-photon entanglement, \href{https://doi.org/10.1038/NPHOTON.2011.354}{Nat. Photonics {\bf 6}, 225 (2012)}.
%
%
\bibitem{expecs3}C. Wang, Y. Y. Gao, P. Reinhold, R. W. Heeres, N. Ofek,
K. Chou, C. Axline, M. Reagor, J. Blumoff, K. M. Sliwa,
L. Frunzio, S. M. Girvin, L. Jiang, M. Mirrahimi,
M. H. Devoret, R. J. Schoelkopf, A Schr${\rm \ddot{o}}$dinger cat living in two boxes, \href{https://doi.org/10.1126/science.aaf2941}{Science {\bf 352}, 1087 (2016)}.
%
%
\bibitem{expecs4}B. Hacker, S. Welte, S. Daiss, A. Shaukat, S. Ritter, L. Li, and
G. Rempe, Deterministic creation of entangled atom-light
Schr${\rm \ddot{o}}$dinger-cat states, \href{https://doi.org/10.1038/s41566-018-0339-5}{Nat. Photon. {\bf 13}, 110 (2019)}.
%
%
\bibitem{expecs5}L. Duan, Creating Schr\"odinger-cat states, \href{https://doi.org/10.1038/s41566-018-0340-z}{Nat. photon. {\bf 13}, 73 (2019)}.
%
%
\bibitem{expecs6}A. Omran, H. Levine, A. Keesling, G. Semeghini, T. T. Wang, S. Ebadi,
H. Bernien, A. S. Zibrov, H. Pichler, S. Choi, J. Cui, M. Rossignolo, P. Rembold , S. Montangero, T. Calarco, M. Endres, M. Greiner, V. Vuleti$\rm\acute{c}$, M. D. Lukin, Generation and manipulation of Schr$\rm{\ddot{o}}$dinger cat states in Rydberg atom arrays, \href{https://doi.org/10.1126/science.aax9743}{Science {\bf 365}, 570 (2019)}.
%
%
\bibitem{DPO1}P. D. Drummond, K. J. McNeil, and D. F. Walls, Non-equilibrium Transitions in Sub/Second
Harmonic Generation, \href{https://doi.org/10.1080/713820226}{Optica Acta {\bf 27}, 321 (1980)};{\it ibid.} \href{https://doi.org/10.1080/713820531}{{\bf 28}, 211 (1981)}.
%
%
\bibitem{DPO2}A. Heidmann, R. J. Horowicz, S. Reynaud, E. Giacobino, C. Fabre, and G. Camy, Observation of Quantum Noise Reduction on Twin Laser Beams, \href{https://doi.org/10.1103/PhysRevLett.59.2555}{Phys. Rev. Lett. {\bf 59}, 2555 (1987)}.
%
%
\bibitem{DPO3}P. Kinsler and P. D. Drummond, Quantum dynamics of the parametric oscillator, \href{https://doi.org/10.1103/PhysRevA.43.6194}{Phys. Rev. A {\bf 43}, 6194 (1991)}.
%
%
\bibitem{DPO4}H. Deng, D. Erenso, R. Vyas, and S. Singh, Entanglement, Interference, and Measurement
in a Degenerate Parametric Oscillator, \href{https://doi.org/10.1103/PhysRevLett.86.2770}{Phys. Rev. Lett. {\bf 86}, 2770 (2001)}.
%

%
\bibitem{CIM1}A. Marandi, Z. Wang, K. Takata, R. L. Byer, and Y. Yamamoto, Network of time-multiplexed optical parametric
oscillators as a coherent Ising machine, \href{https://doi.org/10.1038/NPHOTON.2014.249}{Nat. Photonics {\bf 8}, 937 (2014)}.
%
%
\bibitem{CIM2}K. Takata, A. Marandi, and Y. Yamamoto, Quantum correlation in degenerate optical parametric oscillators with mutual injections, \href{https://doi.org/10.1103/PhysRevA.92.043821}{Phys. Rev. A {\bf 92}, 043821 (2015)}.
%
%
\bibitem{CIM3}T. Inagaki, Y. Haribara, K. Igarashi, T. Sonobe, S. Tamate, T. Honjo, A. Marandi, P. L. McMahon, T. Umeki, K. Enbutsu, O. Tadanaga, H. Takenouchi, K. Aihara, K. Kawarabayashi, K. Inoue, S. Utsunomiya, and H. Takesue, A coherent Ising machine for
2000-node optimization problems, \href{https://doi.org/10.1126/science.aah4243}{Science {\bf 354}, 603 (2016)}.
%
%
\bibitem{CIM4}P. L. McMahon, A. Marandi, Y. Haribara, R. Hamerly, C. Langrock, S. Tamate, T. Inagaki, H. Takesue, S. Utsunomiya, K. Aihara, R. L. Byer, M. M. Fejer, H. Mabuchi, Y. Yamamoto, A fully programmable 100-spin coherent Ising machine with all-to-all connections, \href{https://doi.org/10.1126/science.aah5178}{Science {\bf 354}, 614 (2016)}.
%
%
\bibitem{CIM5}Y. Yamamoto, K. Aihara, T. Leleu, K. Kawarabayashi, S. Kako, M. Fejer, K. Inoue, and
H. Takesue, Coherent Ising machines-optical neural networks operating
at the quantum limit, \href{https://doi.org/10.1038/s41534-017-0048-9}{npj Quantum infromation {\bf 3}, 49 (2017)}.
%
%
\bibitem{CIM6}A. Yamamura, K. Aihara, and Y. Yamamoto, Quantum model for coherent Ising machines: Discrete-time measurement feedback formulation, \href{https://doi.org/10.1103/PhysRevA.96.053834}{Phys. Rev. A {\bf 96}, 053834 (2017)}.
%
%
\bibitem{CIM7}M. Babaeian, D. T. Nguyen, V. Demir, M. Akbulut, P.-A Blanche,
Y. Kaneda, S. Guha, M. A. Neifeld, and N. Peyghambarian, A single shot coherent Ising machine based on a
network of injection-locked multicore fiber lasers, \href{https://doi.org/10.1038/s41467-019-11548-4}{Nat. Commun. {\bf 10}, 3516 (2019)}.
%
%
\bibitem{CIM8}Y. Inui, and Y. Yamamoto, Entanglement and quantum discord in optically coupled coherent Ising machines, \href{https://doi.org/10.1103/PhysRevA.102.062419}{Phys. Rev. A {\bf 102}, 062419 (2020)}.
%
%
\bibitem{CIM9}R. Y. Teh, P. D. Drummond, and M. D. Reid, Overcoming decoherence of Schr\"odinger cat states formed in a cavity using squeezed-state inputs, \href{https://doi.org/10.1103/PhysRevResearch.2.043387}{Phys. Rev. Research {\bf 2}, 043387 (2020)}.
%
%
\bibitem{cat1}C. Monroe, D. M. Meekhof, B. E. King, D. J. Wineland, A ``Schr$\rm \ddot{o}$dinger Cat"
Superposition State of an Atom, \href{https://doi.org/10.1126/science.272.5265.1131}{Science {\bf 272}, 1131 (1996)}.
%
%
\bibitem{cat2}G. S. Agarwal, R. R. Puri, and R. P. Singh, Atomic Schr\"odinger cat states
 ,\href{https://doi.org/10.1103/PhysRevA.56.2249}{Phys. Rev. A {\bf 56}, 2249 (1997)}.
%

%
\bibitem{cat3}D. Leibfried, E. Knill, S. Seidelin, J. Britton, R. B. Blakestad, J. Chiaverini, D. B. Hume, W. M. Itano,
J. D. Jost, C. Langer, R. Ozeri, R. Reichle, and D. J. Wineland, Creation of a six-atom 'Schr\"odinger cat' state, \href{https://doi.org/10.1038/nature04251}{Nature {\bf 438}, 639 (2005)}.
%

%
\bibitem{cat4}B. He, M. Nadeem, and J. A. Bergou, Scheme for generating coherent-state superpositions with realistic cross-Kerr nonlinearity, \href{https://doi.org/10.1103/PhysRevA.79.035802}{Phys. Rev. A {\bf 79}, 035802 (2009)}.
%
%
\bibitem{cat5}$\rm T.\ Opatrn{\acute{y}}\ and\ K.\ M{\o}lmer$, Spin squeezing and Schr\"odinger-cat-state generation in atomic samples with Rydberg blockade, \href{https://doi.org/10.1103/PhysRevA.86.023845}{Phys. Rev. A {\bf 86}, 023845 (2012)}.
%
%
\bibitem{cat6}H. Tan, F. Bariani, G. Li, and P. Meystre, Generation of macroscopic quantum superpositions of optomechanical oscillators by dissipation, \href{https://doi.org/10.1103/PhysRevA.88.023817}{Phys. Rev. A {\bf 88}, 023817 (2013)}.
%
%
\bibitem{cat7}B. Vlastakis, G. Kirchmair, Z. Leghtas, S. E. Nigg, L. Frunzio,
S. M. Girvin, M. Mirrahimi, M. H. Devoret, R. J. Schoelkopf, Deterministically Encoding Quantum
Information Using 100-Photon Schr\"odinger Cat States, \href{https://doi.org/10.1126/science.1243289}{Science {\bf 342}, 607 (2013)}.
%
%
\bibitem{cat8}X. Wang, A. Miranowicz, H.-R. Li, and F. Nori, Hybrid quantum device with a carbon nanotube and a flux qubit for dissipative quantum engineering, \href{https://doi.org/10.1103/PhysRevB.95.205415}{Phys. Rev. B {\bf 95}, 205415 (2017)}.
%

%
\bibitem{dpocat1}F. Minganti, N. Bartolo, J. Lolli, W. Casteels, and C. Ciuti, Exact results for Schr$\rm\ddot{o}$dinger cats in driven-dissipative systems and their feedback control, \href{https://doi.org/10.1038/srep26987}{Sci. Rep. {\bf 6}, 26987 (2016)}.
%
%
\bibitem{dpocat2}S. E. Nigg, N. L$\rm \ddot{o}$rch, and R. P. Tiwari, Robust quantum optimizer with full connectivity, \href{https://doi.org/10.1126/sciadv.1602273}{Sci. adv. {\bf 3}, e1602273 (2017)}.
%
%
\bibitem{dpocat3}S. Puri, S. Boutin, and A. Blais, Engineering the quantum states of light in a Kerr-nonlinear
resonator by two-photon driving, \href{https://doi.org/10.1038/s41534-017-0019-1}{npj Quantum information {\bf 3}, 18 (2017)}.
%
%
\bibitem{dpocat4}C. N. Gagatsos and S. Guha, Efficient representation of Gaussian states for multimode non-Gaussian quantum state engineering via subtraction of arbitrary number of photons, \href{https://doi.org/10.1103/PhysRevA.99.053816}{Phys. Rev. A {\bf 99}, 053816 (2019)}.
%
%
\bibitem{dpocat5}Z. Wang, M. Pechal, E. A. Wollack, P. Arrangoiz-Arriola,
M. Gao, N. R. Lee, and A. H. Safavi-Naeini, Quantum Dynamics of a Few-Photon Parametric Oscillator, \href{https://doi.org/10.1103/PhysRevX.9.021049}{Phys. Rev. X {\bf 9}, 021049 (2019)}.
%
%
\bibitem{dpocat6}R. Y. Teh, F.-X. Sun, R. E. S. Polkinghorne, Q. Y. He, Q. Gong, P. D. Drummond, and M. D. Reid, Dynamics of transient cat states in degenerate parametric oscillation with and without nonlinear
Kerr interactions, \href{https://doi.org/10.1103/PhysRevA.101.043807}{Phys. Rev. A {\bf 101}, 043807 (2020)}.
%
%
\bibitem{dpocat7}W. Qin, A. Miranowicz, H. Jing, and F. Nori, Generating Large Cats with Nine Lives: Long-Lived Macroscopic Schr\"odinger Cat States in Atomic Ensembles, \href{https://arxiv.org/abs/2101.03662}{arXiv:2101.03662 (2021)}.
%
%
\bibitem{mec1}Y. Aharonov, H. Pendleton, and A. Petersen, Modular variables in Quantum theory, \href{https://doi.org/10.1007/BF00670008}{Int. J. Theor. Phys. {\bf 2}, 213 (1969)}.
%
%
\bibitem{mec2}C. Gneiting and K. Hornberger, Detecting entanglement in spatial interference, \href{https://doi.org/10.1103/PhysRevLett.106.210501}{Phys. Rev. Lett. {\bf 106}, 210501 (2011)}.
%
%
\bibitem{mec3}M. A. D. Carvalho, J. Ferraz, G. F. Borges, P.-L de Assis, S. P$\rm \acute{a}$dua, and S. P. Walborn, Experimental observation of quantum correlations in modular variables, \href{https://doi.org/10.1103/PhysRevA.86.032332}{Phys. Rev. A {\bf 86}, 032332 (2012)}.
%
%
\bibitem{mec4}C. Gneiting and K. Hornberger, Nonlocal Young tests with Einstein-Podolsky-Rosen-correlated particle pairs, \href{https://doi.org/10.1103/PhysRevA.88.013610}{Phys. Rev. A {\bf 88}, 013610 (2013)}.
%
%
\bibitem{mec5}J. C. G. Biniok, P. Busch, and J. Kiukas, Uncertainty in the context of multislit interferometry, \href{https://doi.org/10.1103/PhysRevA.90.022115}{Phys. Rev. A {\bf 90}, 022115 (2014)}.
%
%
\bibitem{dphotondecay1}H. D. Simaan and R. Loudon, Off-diagonal density matrix for single-beam twophoton
absorbed light, \href{https://doi.org/10.1088/0305-4470/11/2/018}{J. Phys. A: Math. Gen. {\rm 11}, 435 (1978)}.
%
%
\bibitem{dphotondecay2}L. Gilles and P. L. Knight, Two-photon absorption and nonclassical states of light, \href{https://doi.org/10.1103/PhysRevA.48.1582}{Phys. Rev. A {\bf 48}, 1582 (1993)}.
%
%
\bibitem{dphotondecay3}L. Gilles, B. M. Garraway, and P. L. Knight, Generation of nonclassical light by dissipative two-photon processes, \href{https://doi.org/10.1103/PhysRevA.49.2785}{Phys. Rev. A {\bf 49}, 2785 (1994)}.
%
%
\bibitem{dphotondecay4}V. V. Dodonov and S. S. Mizrahi, Competition between one- and two-photon absorption
processes, \href{https://doi.org/10.1088/0305-4470/30/9/008}{J. Phys. A: Math. Gen. {\bf 30}, 2915 (1996)}.
%
%
\bibitem{dphotondecay5}E. S. Guerra, B. M. Garraway, and P. L. Knight, Two-photon parametric pumping versus two-photon absorption: A quantum jump approach, \href{https://doi.org/10.1103/PhysRevA.55.3842}{Phys. Rev. A {\bf 55}, 3842 (1997)}.
%
%
\bibitem{dphotondecay6}M. Mirrahimi, Z. Leghtas, V. V Albert, S. Touzard, R. J Schoelkopf, Liang Jiang, and M. H Devoret, Dynamically protected cat-qubits: a new paradigm for universal quantum computation, \href{https://doi.org/10.1088/1367-2630/16/4/045014}{New J. Phys. {\bf 16}, 045014 (2014)}.
%
%
\bibitem{dphotondecay7}A. Miranowicz, J. Bajer, M. Paprzycka, Y.-X. Liu, A. M. Zagoskin, and F. Nori, State-dependent photon blockade via quantum-reservoir engineering, \href{https://doi.org/10.1103/PhysRevA.90.033831}{Phys. Rev. A {\bf 90}, 033831 (2014)}.
%
%
\bibitem{CIMstability}M. C. Strinati, L. Bello, E. G. D. Torre, and A. Pe'er, Can nonlinear parametric oscillators solve random ising models?, \href{https://doi.org/10.1103/PhysRevLett.126.143901}{Phys. Rev. Lett. {\bf 126}, 143901 (2021)}.
%
%
\bibitem{gse1}L.-M. Duan, G. Giedke, J. I. Cirac, and P. Zoller, Inseparability Criterion for Continuous Variable Systems, \href{https://doi.org/10.1103/PhysRevLett.84.2722}{Phys. Rev. Lett. {\bf 84}, 2722 (2000)}.
%
%
\bibitem{gse2}R. Simon, Peres-Horodecki Separability Criterion for Continuous Variable Systems, \href{https://doi.org/10.1103/PhysRevLett.84.2726}{Phys. Rev. Lett. {\bf 84}, 2726 (2000)}.
%
%
\bibitem{Wignerfunction1}W. F. Braasch, J.O. D. Friedman, A. J. Rimberg, and M. P. Blencowe, Wigner current for open quantum systems, \href{https://doi.org/10.1103/PhysRevA.100.012124}{Phys. Rev. A {\bf 100}, 012124 (2019)}.
%
%
\bibitem{sqb1}W. Qin, A. Miranowicz, P.-B. Li X.-Y. L$\rm \ddot{u}$, J. Q. You, and F. Nori, Exponentially enhanced light-matter interaction, cooperativities, and steady-state entanglement using parametric amplification, \href{https://doi.org/10.1103/PhysRevLett.120.093601}{Phys. Rev. Lett. {\bf 120}, 093601 (2018)}.
%
%
\bibitem{disspativestabilization}M. Mamaev, L. C. G. Govia, and A. A. Clerk, Dissipative stabilization of entangled cat states using a driven Bose-Hubbard dimer, \href{https://doi.org/10.22331/q-2018-03-27-58}{Quantum {\bf 2}, 58 (2018)}.
%
%
\bibitem{sqb2}Y.-H. Chen, W. Qin, X. Wang, A. Miranowicz, and F. Nori, Shortcuts to Adiabaticity for the Quantum Rabi Model: Efficient Generation of Giant Entangled Cat States via Parametric Amplification, \href{https://doi.org/10.1103/PhysRevLett.126.023602}{Phys. Rev. Lett. {\bf 126}, 023602 (2021)}.
%
\end{references}
\end{document}